\documentclass[onecolumn]{revtex4}
\pdfoutput=1
\usepackage{amsmath}
\usepackage{graphicx}
\usepackage{fancyhdr}
\usepackage{overpic}
\usepackage{float}

\begin{document}
\pagestyle{fancy}

\title{Natural Inflation with a periodic non-minimal coupling}

\rhead{}

\author{Ricardo Z. Ferreira$^{1}$}
	\email{rferreira@icc.ub.edu}
	\author{Alessio Notari$^{1}$}
	\email{notari@fqa.ub.edu}
	\author{Guillem Simeon$^{1}$}
	\email{guillem.simeon@gmail.com}
	
	\affiliation{$^{1}$ Departament de F\'isica Qu\`antica i Astrofis\'ica \& Institut de Ci\`encies del Cosmos (ICCUB), Universitat de Barcelona, Mart\'i i Franqu\`es 1, 08028 Barcelona, Spain}

\begin{abstract}
Natural inflation is an attractive model for primordial inflation, since the potential for the inflaton is of the pseudo Nambu-Goldstone form, $V(\phi)=\Lambda^4 [1+\cos (\phi/f)]$, and so is protected against radiative corrections. Successful inflation can be achieved if $f \gtrsim {\rm few}\, M_{P}$ and $\Lambda \sim m_{GUT}$ where $\Lambda$ can be seen as the strong coupling scale of a given non-abelian gauge group. However, the latest observational constraints put natural inflation in some tension with data. We show here that a non-minimal coupling to gravity $\gamma^2(\phi) R$, that respects the symmetry $\phi\rightarrow \phi+2 \pi f$ and has a simple form, proportional to the potential, can improve the agreement with cosmological data. Moreover, in certain cases, satisfactory agreement with the Planck 2018 TT, TE, EE and low P data can be achieved even for a periodicity scale of approximately $M_p$.
\end{abstract}

\maketitle

\section{ Introduction}

Inflationary models where the inflaton is a pseudo-Goldstone boson can naturally provide a flat potential, which is protected against radiative corrections. 
Natural Inflation (NI) \cite{Freese:1990rb} realizes this idea using a continuous shift symmetry of an axion-like field broken down to a discrete shift symmetry by non-perturbative effects associated with a non-abelian gauge field that becomes strongly coupled at a scale $\Lambda$, thus, generating a potential of the form $V(\phi)=\Lambda^4 [1+\cos (\phi/f)]$.

In this letter we consider a minimal extension of the original NI model by considering the simplest non-minimal coupling of the inflaton to gravity $\gamma^2(\phi) R$, which respects the discrete shift symmetry $\phi\rightarrow \phi+2 \pi f$. We require Einstein gravity to be recovered at the minimum of the potential and we also assume the non-minimal coupling function to be proportional to $V(\phi)$. This last requirement encodes the fact that gravity feels the field value $\phi$ through its potential energy, and so the process generating such non-minimal coupling should have that information. In other words, any coupling to $R$ which is non-derivative in $\phi$ should vanish if $V(\phi)$ goes to zero. A similar logic of adding shift symmetric corrections to NI was also proposed in \cite{Germani:2010hd}, using only derivative interactions, and in~\cite{Alvarez:2017kar}, using non-standard kinetic terms.

The motivation to consider this extension is three-fold. First, from an EFT point of view all the couplings allowed by the symmetries should be considered. Second, and on a more practical level, the predictions of natural inflation are in tension with the latest Planck 2018 results \cite{Akrami:2018odb} and so it is interesting to understand what kind of extensions could alleviate the tension. Third, in standard NI the axion decay constant $f$ needs to be super-Planckian in order to fit cosmological data. This feature might be problematic due to the presence of  gravitational instanton corrections \cite{Banks:2003sx,Rudelius:2015xta,Rudelius:2014wla,Montero:2015ofa,Hebecker:2016dsw}. 
Several extensions of the original NI model have been proposed to overcome one or both of the last two issues (for an incomplete list see \cite{ArkaniHamed:2003wu, Kim:2004rp, Dimopoulos:2005ac, Anber:2009ua, Bachlechner:2014hsa, Kaloper:2014zba, DAmico:2017cda, Ferreira:2017lnd,Ferreira:2017wlx}), although some of the proposals still seem to inherit the gravitational instanton problem~\cite{Rudelius:2014wla,Rudelius:2015xta,Montero:2015ofa,Hebecker:2016dsw}.

As we will show, the simple extension we consider here comes with a set of interesting predictions which can address both issues.

\section{Non-minimal coupling to gravity}

We consider natural inflation with a non-minimal coupling to gravity described by the action:
\begin{equation}
S=\int d^{4}x \sqrt{|g|} \bigg[\frac{1}{2}{M_{P}^{2}} \gamma(\phi)^2 R-\frac{1}{2} {g}^{\mu\nu} \partial_{\nu}\phi \partial_{\nu}\phi -V(\phi) \bigg] \, ,
\end{equation}
where $V(\phi) = \Lambda^{4} \Big[1+\cos{(\phi/f)} \Big]$. In standard NI \cite{Freese:1990rb} $\Lambda$ turns out to be of the order of the GUT scale and the periodicity scale $f$ has to take values larger than the (reduced) Planck mass, $M_P\simeq 2.4\times 10^{18}$GeV, in order to achieve successful inflation, i.e., in order to drive 60 efolds of slow-roll inflation. Here we add the simplest non-minimal coupling compatible with the periodicity of the original potential, that gives standard Einstein gravity at the minimum of the potential and  such that the deviation is proportional to the potential of $\phi$, 
\begin{equation}
\gamma(\phi)^2 \equiv 1+ \alpha \Big[ 1+\cos{\Big(\frac{\phi}{f} \Big)} \Big] \, ,
\end{equation}
where the dimensionless number $\alpha$ is the only new parameter. Note also that $\alpha > -\frac{1}{2}$ for the Planck scale ${M_{P} \gamma}$ to be well-defined.

The usual way to get rid of the non-minimal coupling from the action is to change variables, going from the so-called Jordan frame to the Einstein frame, by means of a conformal transformation of the metric:
\begin{equation}
\tilde{g}_{\mu\nu} = {\gamma}^2 g_{\mu\nu} \label{conf} \, .
\end{equation}

Such a transformation leads to a non-canonical contribution to the kinetic term and a rescaling of the potential. Explicitly, the action in the Einstein frame reads:
\begin{equation}
\begin{gathered}
S=\int d^{4}x \sqrt{-\tilde{g}} \bigg(\frac{1}{2}{M_{P}^{2}} \tilde{R} -\frac{1}{2} K(\phi) {\tilde{g}}^{\mu\nu}{\partial_{\mu}}\phi{\partial_{\nu}}\phi -\frac{V(\phi)}{\gamma^4} \bigg) \, ,
\end{gathered}
\end{equation}
 where quantities computed with the transformed metric are denoted with a tilde and 
 \begin{eqnarray}
 K(\phi) \equiv \frac{1 +6 M_P^2 \gamma'^2}{\gamma^2} = \frac{2 {\gamma}^2 f^2 + 3 {M_{P}^{2}} {\alpha}^2 {\sin^2{\frac{\phi}{f}}}}{2 \gamma^4 f^2}, \label{Kphi}
 \end{eqnarray}
 where a prime denotes $d/d\phi$. By redefining the scalar field using the transformation 
\begin{equation}
\frac{d\chi}{d\phi} = \sqrt{K(\phi)} \, , \label{chiphi}
\end{equation}
we obtain the action in the Einstein frame in terms of a new canonical field $\chi$ with an effective potential $U(\chi) \equiv V(\phi(\chi))/\gamma(\phi(\chi))^4$:
\begin{equation}
S=\int d^{4}x \sqrt{-\tilde{g}} \bigg(\frac{1}{2}{M_{P}^{2}} \tilde{R}-\frac{1}{2}{\tilde{g}}^{\mu\nu}{\partial_{\mu}}\chi{\partial_{\nu}}\chi -U(\chi) \bigg) \, ,
\end{equation}
making the standard slow-roll analysis possible.

\begin{figure}
	\begin{center}
		\vspace*{3mm}
		\includegraphics[scale=0.4]{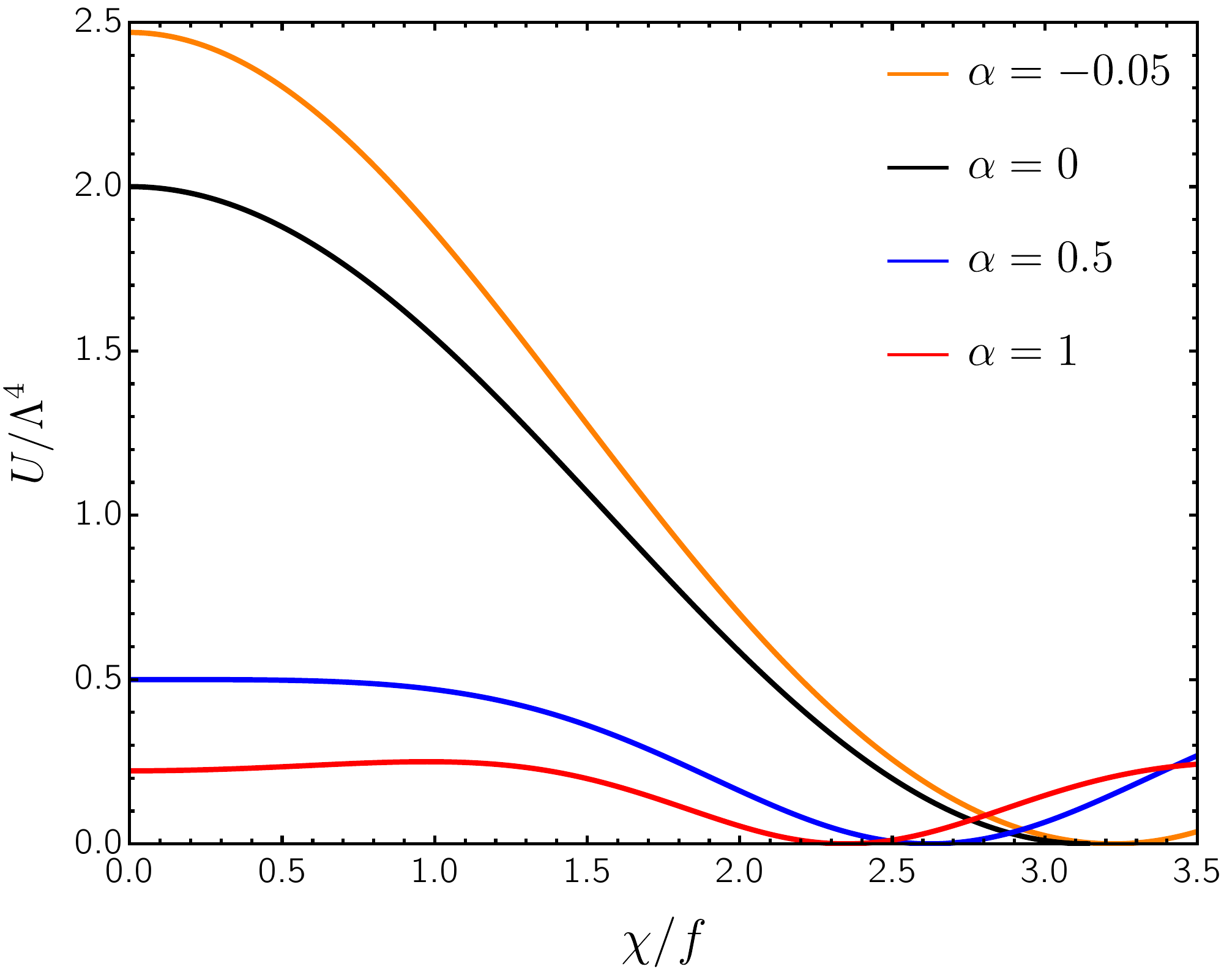}
		\caption {Normalized effective potential $U/\Lambda^4$ in the Einstein frame as a function of $\chi/f$.\label{fig1}     }
	\end{center}
\end{figure}

\section{Slow-roll analysis}
\begin{figure}
	\begin{center}
		\vspace*{3mm}
		\includegraphics[scale=0.4]{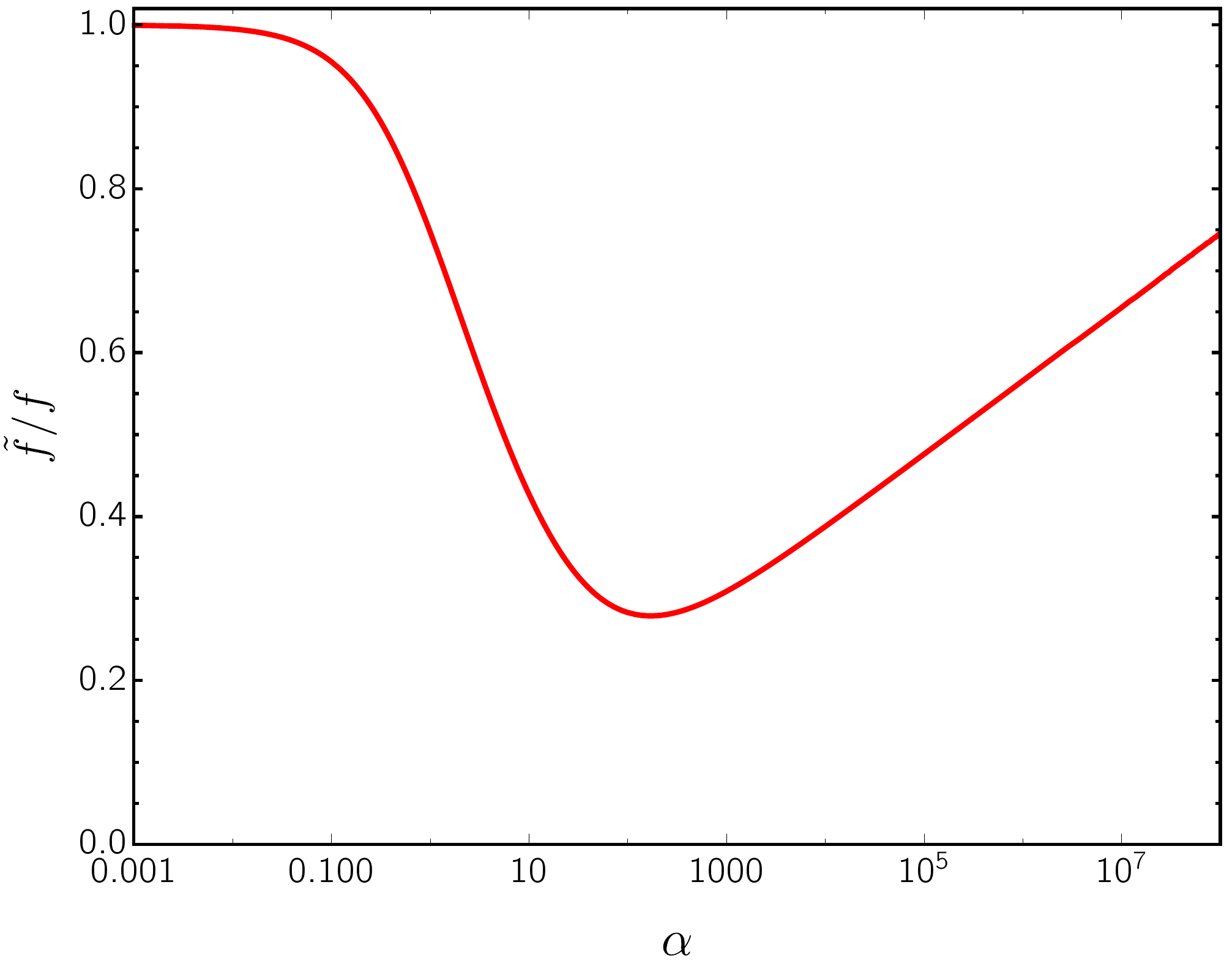}
		\caption{Ratio between the  periodicity scale $\tilde{f}$, in the Einstein frame and with canonical field $\chi$, vs. the periodicity scale $f$ in the Jordan frame potential $V(\phi)$, as a function of $\alpha$. \label{ftilde}}
	\end{center}
\end{figure}

The potential in the Einstein frame $U(\chi)$, shown in fig.~\ref{fig1},  flattens as $\alpha$ approaches 0.5, while new extrema appear for $\alpha > 0.5$. The potential $U(\chi)$ is still periodic, but with a new periodicity scale, $\tilde{f}$, which turns out to be different from the scale $f$, present in the original potential, although still of the same order of magnitude. In terms of the canonically normalized field $\chi$ the computation of the instanton action reduces to what was derived in \cite{Giddings:1987cg, Montero:2015ofa} with a periodicity scale $\tilde{f}$. In fig.~\ref{ftilde} we show the full dependence of $\tilde{f}/f$ on $\alpha$. Actually, for $\alpha \lesssim 1$, which is one of the interesting regions for this scenario as we will explain below, the two scales are very close together.
When comparing the periodicity scale to the observed Planck scale we will always refer to their ratio $\tilde{f}/M_P$ in the Einstein frame, where the axion is canonical and the Planck mass is a constant. 
In this frame the Friedmann equation and the axion equation of motion yield simply
\begin{eqnarray}
3M_{P}^{2}H^2 &=& U + \frac{\dot{\chi}^2}{2} \, \label{FRW} \nonumber \, , \\
\ddot{\chi} + 3H\dot{\chi}+\frac{dU}{d\chi} &=& 0 \, . \label{KG}
\end{eqnarray}
Inflation starts when the slow-roll conditions, {\it i.e.} $\epsilon \ll 1$ and $|\eta| \ll 1$, are satisfied, where
\begin{eqnarray}
\epsilon  \equiv  \frac{{M_{P}^{2}}}{2} \frac{1}{U^2} {\bigg(\frac{dU}{d\chi}\bigg)}^2 \, , \qquad \,
\eta \equiv  {M_{P}^{2}} \frac{1}{U} {\bigg(\frac{d^{2}U}{d\chi^2}\bigg)} \, ,
\end{eqnarray}
and it ends when $\epsilon \simeq 1$, which determines the field value at the end of inflation, $\chi_{end}$.
The number of e-foldings $ N$ of inflation from a given value $\chi_{0}$ to $\chi_{end}$ is given by:
\begin{equation}
 N = - \frac{1}{{M_{P}^{2}}} \int_{\chi_0}^{\chi_{end}} U {\bigg(\frac{dU}{d\chi}\bigg)}^{-1} d\chi \label{Nefolds} \, .
\end{equation}
The observable scales correspond to $N=\Delta N$ e-folds before the end of inflation. 
We will use $55\lesssim \Delta N \lesssim 62$, the uncertainty being, as usual, due to model-dependence in the reheating history. The observational constraint $P_\zeta = {H}^{4}/(2\pi \dot{\chi})^2|_{\Delta N}= 2.2 \times 10^{-9}
$ imposes a relation between the values of $\Lambda$, $\alpha$ and $f$. We use this constraint to eliminate the scale $\Lambda$, and we vary the two independent parameters $\alpha$ and $f$. We will, however, present the results as a function of $\alpha$ and $\tilde{f}$.
Finally, we compute the scalar spectral index $n_{s} \approx (1 - 6\epsilon + 2\eta)|_{\Delta N} $ and the tensor-to-scalar ratio $r \approx 16\epsilon|_{\Delta N}$.

\section{Results}
\begin{figure}
\begin{center}
\vspace*{3mm}
\includegraphics[scale=0.43]{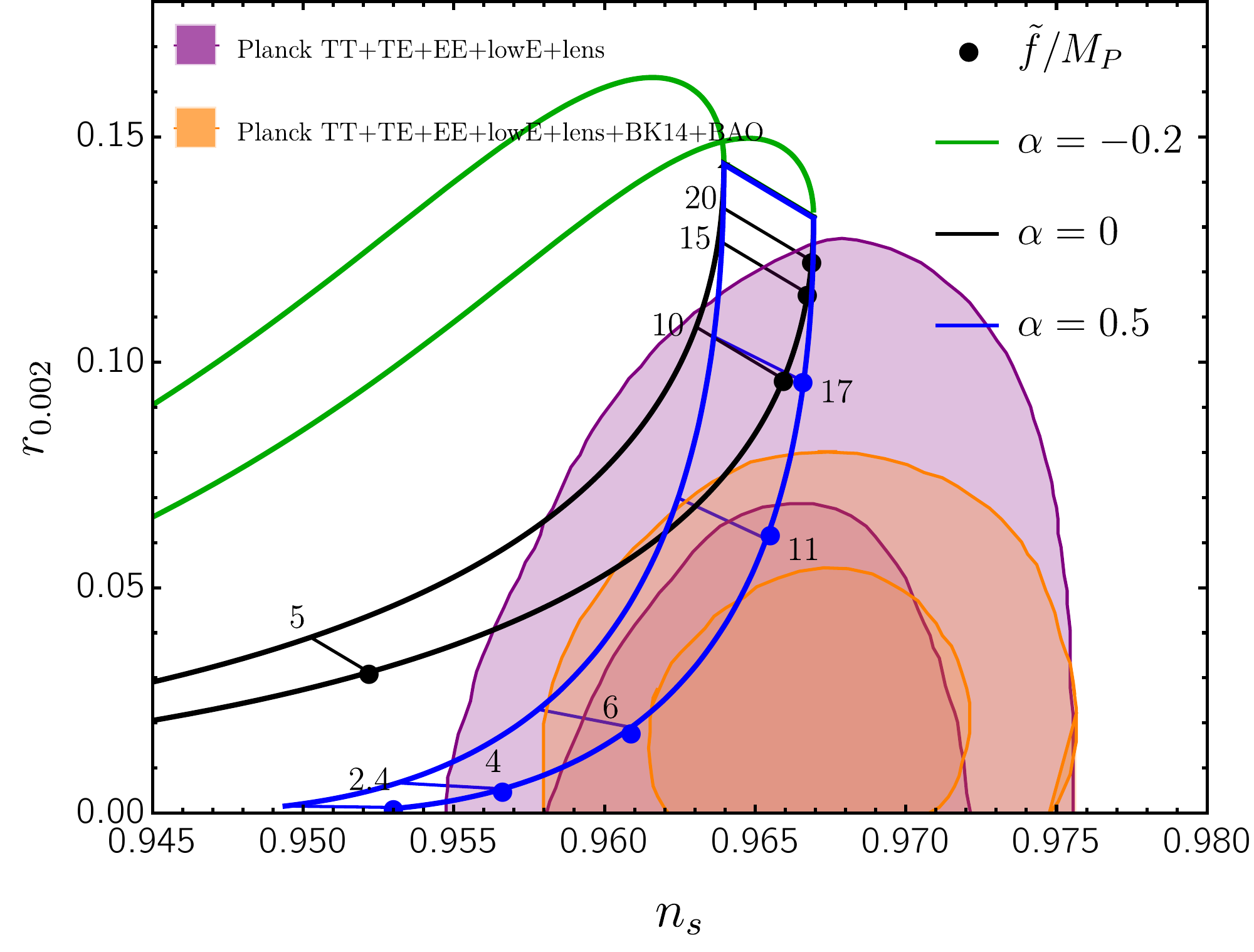} 
\includegraphics[scale=0.43]{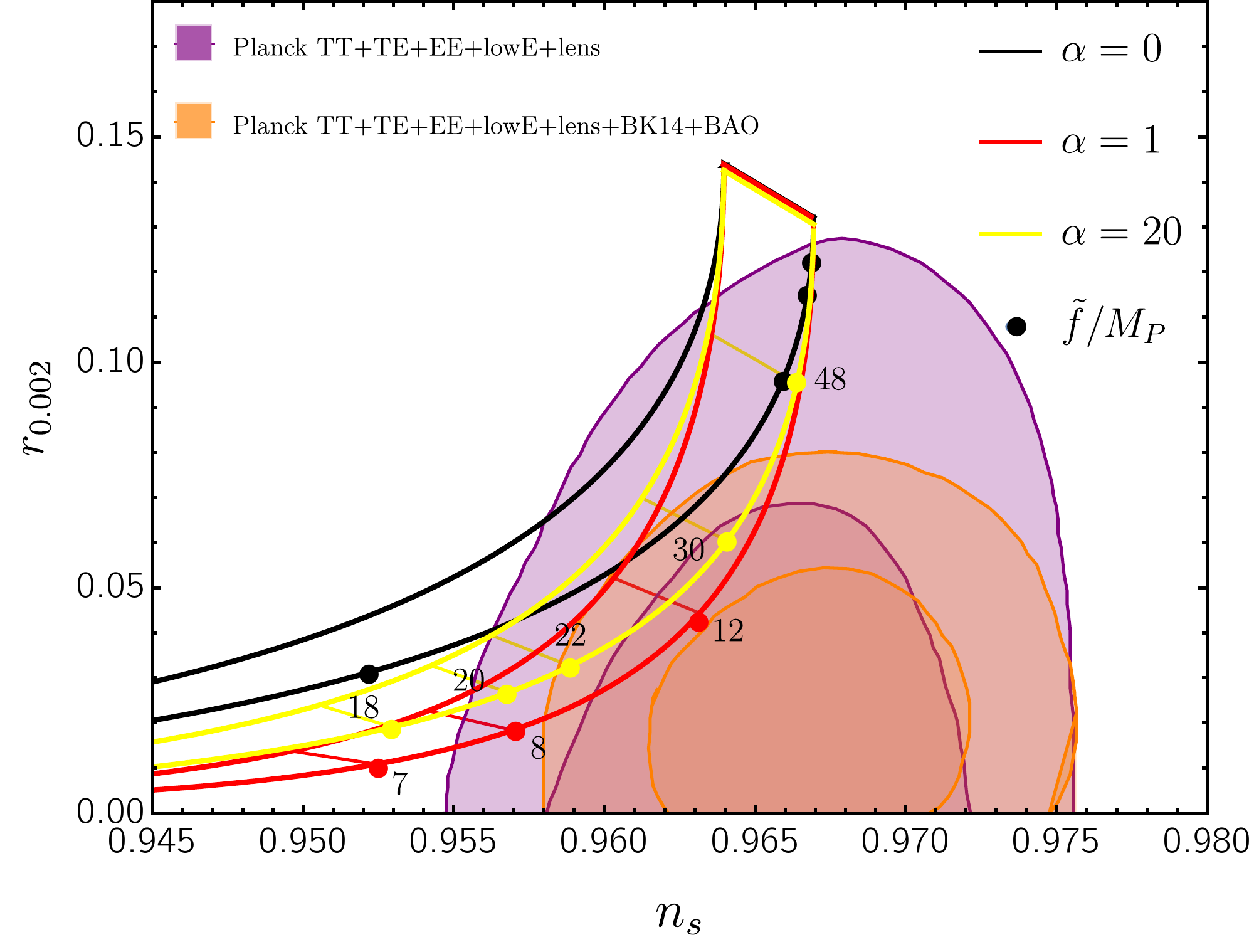}
\caption{Scalar spectral index $n_{s}$ and  tensor-to-scalar ratio $r_{0.002}$, obtained varying $\alpha$ and $\tilde{f}$. There are two lines per color, corresponding to $\Delta N= 55$ and $\Delta N=60$. The shaded purple regions are the observational constraints from Planck TT, TE, EE and low P data while the orange ones are the constraints from Planck TT and low P data combined with Bicep-Keck and BAO data sets \cite{Akrami:2018odb}. The darker regions represent 68\% C.L. and the lighter ones 95\% C.L.. \label{fig2}}
\end{center}
\end{figure}

\begin{figure}
\begin{center}
\vspace*{3mm}
\includegraphics[scale=0.43]{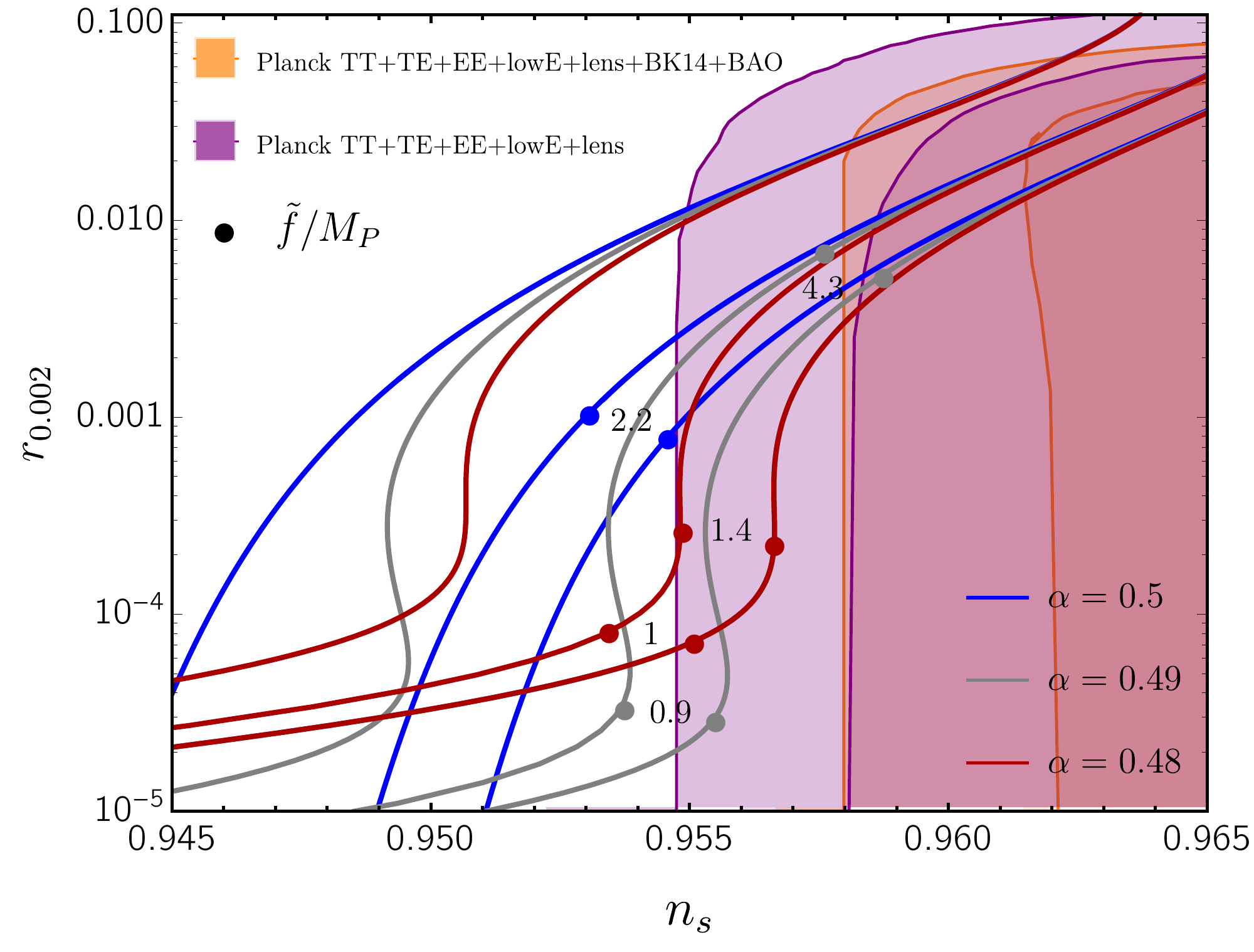}
\caption {Results obtained by running $f$ from approximately $0.5 M_{P}$ to $10 M_{P}$ for $\Delta N=55$ (left), $\Delta N=60$ (middle) and $\Delta N=62$ (right). The shaded purple regions are the observational constraints from Planck TT, TE, EE and low P data while the orange ones are the constraints from Planck TT and low P data combined with Bicep-Keck and BAO data sets \cite{Akrami:2018odb}. The darker regions represent 68\% C.L. and the lighter ones 95\% C.L..  \label{fig3}}
\end{center}
\end{figure}

In figs.~\ref{fig2} and \ref{fig3} we show the results for $n_s$ and $r$ for different values of $\alpha, \tilde{f}/M_P$ and $\Delta N$ after solving numerically eqs.~(\ref{KG}-\ref{Nefolds}). In the same figures we also show the 68\% and 95\% C.L.  contours coming from the 2018 Planck TT, TE, EE, low E and lensing data and also combining with BAO and Bicep-Keck data (BK14)~\cite{Akrami:2018odb}. Unless otherwise stated, we will take the former constraints as a benchmark for the rest of the analysis.

The results show that $\alpha > 0$ suppresses the amount of tensor modes with respect to standard NI. This is expected since increasing $\alpha$ lowers and flattens the potential, as can be seen from fig.~\ref{fig1}. This suppression alleviates some of the current tension of NI with the observational constraints: the values for $n_{s}$ and $r_{0.002}$ with a non-minimal coupling to gravity are found to be well within the 95\% C.L. region for a wide range of parameters and, for $\Delta N \gtrsim 60$, the values reach the 68\% C.L. region.  On the other hand, negative values of $\alpha$ worsen the compatibility with observations, compared to the ones predicted by NI, with predictions excluded from the 95\% C.L. region. In all further analysis, this case will be omitted.

For $\Delta N=60$ we find, for instance, that: $\alpha=0.5$ gives results in the 68\% C.L. region, when $5 M_P \lesssim \tilde{f} \lesssim 12 M_P$; for $\alpha=1$ this happens for $9 M_P \lesssim \tilde{f} \lesssim 15 M_P$; and in the case of $\alpha=20$ the region is reached when $24 \lesssim \tilde{f}/M_P \lesssim 31 M_P$. Values of $\alpha$ larger than 20 show a saturation and give roughly the same curve in the $n_s-r$ plane, although with different values of $f$ (or $\tilde{f}$). The reason is that if we increase $\alpha$ for fixed $f$ the potential in the Einstein frame becomes proportional to $\exp(-\sqrt{2/3} \,\chi/M_p)$, as can be seen from eq.~(\ref{Kphi}),  and no longer supports inflation. Therefore, $f$ also needs to increase. On the other hand for large $f$ the relevant region for observations is the one close to the bottom of the potential. Thus, if we expand around the minimum, in the Jordan frame, both the potential and the non-minimal coupling become quadratic in $\phi$, the latter with a strength given by $\alpha/f^2$. From this we can see that in this limit increasing $\alpha$ can be compensated with a larger $f$ to get the same prediction. Generically, but except for $\alpha$ close to 0.5, in order to obtain predictions in agreement with Planck data, the scale $\tilde{f}$ needs to be increased when $\alpha$ is also increased.

The region around $\alpha=0.5$ is special because is when the potential becomes shallower at the top. In this region both the scales $\tilde{f}$ and $f$ can be roughly Planckian. In fig.~\ref{fig3} we show such cases, which are characterized by very small tensors, $ 10^{-5} \lesssim r \lesssim 10^{-3}$, for $0.48 \lesssim  \alpha\lesssim 0.5$. This range of tensor modes is probably unobservable by near future experiments.

For the values of $\alpha$ and $f$ that give spectral indices $n_s$ and tensor-to-scalar ratios $r$ lying inside the 68\% C.L. region of the Planck data (see Fig. 1), $\Lambda \sim 10^{15-16}$ GeV is obtained in accordance to the expected value for the GUT scale. For example, for $\alpha=0.4$ and $\tilde{f} \approx 8 M_{P}$ we obtain $\Lambda \approx 1.8 \times 10^{16}$ GeV, while for $\alpha=0.49$ and $\tilde{f}=0.8M_p$ we obtain $\Lambda=2.4\times 10^{15}$ GeV. However, for values of $\alpha$ larger than those plotted in the figures, in order to be inside the 68\% C.L. contour $\Lambda$ also needs to increase such that $\Lambda^4/\alpha^2$, the scale of the potential in the Einstein frame, remains roughly constant. For example, for $\alpha=10^3$ we get $\Lambda\simeq 3.3 \times 10^{17}$ GeV.

When considering the constraint from the combination of the Planck 2018 data with BAO and BK14 data, the model still lies well within the 95\% CL region. The 68\% CL is, instead, only reached for a small region of $\tilde{f}$ for $\alpha$ around 0.5: $\alpha = 0.5$ gives results inside the 68\% CL contour from $\tilde{f} \approx 6 M_P$ to $\tilde{f} \approx 10 M_P$.

One can also estimate the reheating temperature, and so the exact number of efolds $\Delta N$, assuming a given decay process for the inflaton. For example, using the typical perturbative decay of an axion into gauge bosons with a rate $\Gamma=m^3 \alpha^2/(256 \pi^2 f^2)$ we get, assuming that the inflation oscillates until reheating is complete, $T_{RH}\approx 10^8$ GeV and $\Delta N\approx 55$, where we used $f=0.8 M_P$ and $\Lambda=2.4\times 10^{15}$ GeV, a number of relativistic species $g_*=10^2$ and a gauge coupling $\alpha\equiv g^2/(4 \pi)=1$. Note, however, that larger decay rates are expected if the axion decays at tree level into fermions with a large Yukawa coupling~\cite{Ferreira:2017lnd,Ferreira:2017wlx}, or if it decays through a parametric resonance~\cite{Adshead:2015pva,Adshead:2017xll,Notari:2016npn}, leading to a larger $\Delta N$. For instantaneous reheating, in fact, one gets for the same value of $f$ and $\Lambda$, $T_{RH}\approx 10^{15}$ GeV and $\Delta N=60.4$. 

\section{Conclusions}
We have considered in this work a non-minimal coupling between the inflaton and gravity, in the context of natural inflation, that respects the symmetry $\phi\rightarrow \phi+2\pi f$. Assuming a simple cosine form, proportional to the potential itself, with only one extra dimensionless parameter, $\alpha$, we have obtained a model that gives rise to predictions for the spectral index $n_s$ and the tensor-to-scalar ratio $r$ that lay well within the 95\% C.L. region from the combined Planck 2018+BAO+BK14 data.  The parameters that give rise to these results yield a scale $\Lambda \sim 10^{15-16}$ GeV, consistent with the expected value of the GUT scale.

Another interesting consequence of the non-minimal coupling is that inflation can be observationally viable also for values of both $f$ and the periodicity scale in the Einstein frame $\tilde{f}$ closer to $M_p$. Indeed, we have shown that for $0.48 \lesssim \alpha < 0.5$ we can have $\tilde{f}, f \simeq M_P$, within the 95\% C.L. of the Planck 2018 data, although when combining with Bicep-Keck and BAO data one needs $\tilde{f}, f \gtrsim 4 M_p$. This is an improvement over the predictions made by minimally coupled Natural Inflation which only enters the 95\% C.L. of the Planck 2018 data for $f \gtrsim 5.5 M_p$, and of the combined data sets $7.7 \gtrsim f/M_p \gtrsim 6.5$. This case corresponds to smaller values of the tensor-to-scalar ratio $10^{-5} \lesssim r\lesssim 10^{-3}$ and it might alleviate possible issues arising due to gravitational instantons corrections~\cite{Banks:2003sx, Rudelius:2014wla, Montero:2015ofa}, which have been estimated to be exponentially small if the periodicity scale is sub-Planckian.

\begin{acknowledgments}
	We would like to acknowledge Jaume Garriga, Cristiano Germani and Javier Rubio for discussions.
This work is supported by the grants EC FPA2010-20807-C02-02, AGAUR 2009-SGR-168, ERC Starting Grant HoloLHC-306605 and by the Spanish MINECO under MDM-2014-0369 of ICCUB (Unidad de Excelencia ``Maria de Maeztu''). A.N. is grateful  to the Physics Department of the University of Padova for the hospitality during the course of this work and was supported by the ``Visiting Scientist" program of the University of Padova.
\end{acknowledgments}

\bibliography{NIwithNC}

\end{document}